# Secure 3D medical Imaging

**Shadi Al-Zu'bi**
Computer Science Department, AlZaytoonah University of Jordan
Amman, Jordan
Smalzubi@zuj.edu.jo

**Abstract** Image segmentation has proved its importance and plays an important role in various domains such as health systems and satellite-oriented military applications. In this context, accuracy, image quality, and execution time deem to be the major issues to always consider. Although many techniques have been applied, and their experimental results have shown appealing achievements for 2D images in real-time environments, however, there is a lack of works about 3D image segmentation despite its importance in improving segmentation accuracy. Specifically, HMM was used in this domain. However, it suffers from the time complexity, which was updated using different accelerators. As it is important to have efficient 3D image segmentation, we propose in this paper a novel system for partitioning the 3D segmentation process across several distributed machines. The concepts behind distributed multimedia network segmentation were employed to accelerate the segmentation computational time of training Hidden Markov Model (HMMs). Furthermore, a secure transmission has been considered in this distributed environment and various bidirectional multimedia security algorithms have been applied. The contribution of this work lies in providing an efficient and secure algorithm for 3D image segmentation. Through a number of extensive experiments, it was proved that our proposed system is of comparable efficiency to the state of art methods in terms of segmentation accuracy, security and execution time.

Shadi Al-Zu'bi*
Faculty of Science and IT, Al-Zaytoonah University of Jordan, Amman, Jordan
Tel: (+962)799100034
E-mail: smalzubi@zuj.edu.jo

## 1 Introduction

Multimedia information is largely transferred over the Internet due to the low cost and availability. This plays an important role in growing up the use of the Internet in the digital world. The digital image is more appealing than the traditional text in our society. Recently, many interdisciplinary research works have been conducted and proposed in a bid to use computer systems in the process of medical image diagnosis. As an example, detecting an object of interest in a timely manner way. In medical applications, the accuracy of special importance. As the number of health-related problems and challenges is huge, new and emerging technologies can be assumed [23, 24, 26, 28].

Put simply, image segmentation is the process of partitioning a digital image into many image objects. It has many applications in various domains such as content-based image retrieval, computer vision, medicine, traffic con- trol, and security, to name but a few. One of the main applications of image segmentation is in medical image processing (MIP) field. This domain has attracted a lot of works recently due to its importance in the diagnosis of diseases. Specifically, it plays an important role in separating organic tissues that belong to different parts of the human body. The segmentation process can be a 2D segmentation, whereas the output is a two-dimensional segment, or a 3D segmentation, whereas the output is a three-dimensional segment. Obviously, the use of 3D image segmentation in medical diagnosis would increase its accuracy as it would provide a wider view with detailed results to the expert. This would play an important role in eliminating the noise, and as a result, improving the diagnosis accuracy. This was our motivation in this work to use 3D image segmentation.

Although many works have proposed 3D medical image segmentation, which used mathematical morphology and watershed [80, 86, 25, 5, 18], hard or fuzzy clustering, fuzzy connectedness [75], level sets, deformable models, multi-view learning [69, 68, 1, 41, 34], and intelligent scissors [86]. Many other concepts could be used in image segmentation including edge-based segmentation, the use of different layers, and multidimensional images operation [85, 7, 21, 3]. Some of these methods failed to prove their efficiency either in real-time environments or in segmentation accuracy. These real-time environments are essential especially in the remote diagnosis process. On one hand, 3D image segmentation is often a burdensome and computationally expensive operation to perform, as this operation requires processing many 2D images. On the other hand, the real-time environment overwhelms the segmentation method. Therefore, it is vital to find a 3D efficient medical segmentation method in a real-time environment.

The work presented in this paper shows how CAD systems could be employed effectively to segment 3D objects of interest in a real-time environment. This solution is based on distributing the entire medical volume over a multimedia network for distributed execution. Furthermore, transmission over the distributed multimedia network should be secured bidirectionally due to the sensitivity of the transmitted data. Therefore, we adopted the use of Advanced

Encryption Standard (AES) encryption method that divides each image into a multi matrix of pixels and encrypt each matrix to be distributed securely in a different order over the multimedia network, where only the selected distributed devices can decrypt the encrypted matrices to obtain the original image. It should be noted that various window sizes were compared, from which, two were selected for the experiments; (4x4) and (9x9) because of its promising achieved results.

As for the dataset, two datasets were used; a real dataset and two synthesized phantom data-sets. As for the evaluation measurements, we used accuracy, time complexity, encryption errors, and distribution over multimedia network latency. To evaluate the accuracy, we used the human assessment to obtain the accuracy related to the real dataset. As for the synthesized datasets, we use predefined data to get the output. To evaluate the system, we compare it with many existing methods including 3D Geometric HMM, Haar, Iterative Thresholding, kMeans, and many more. In addition, the ability to detect objects based on their diameters was studied, and objects were divided into groups accordingly.

The main contributions of this work are the following.

- Proposing an efficient method to perform real-time 3D image segmentation via distributed devices over multimedia network.
- Adopting security methods to ensure that the process is secure when the data privacy is required.

This work would have many theoretical impacts. In one hand, this work would improve the literature by providing efficient and secure 3D image segmentation method. On the other hand, this work aims at attracting more works for this direction and insisting on the importance of having more stud- ies on secure 3D image segmentation. The practical impacts of this work would lie in further improving the medical diagnosis in hospitals , medical labs, and in the health sector in general. The rest of the paper is structured as follows. Section two reviews the literature and related works in image segmentation, distributed protocols, intelligent multimedia networks, and security methods. Section three provides a detailed description of our proposed method. Sec- tion four details the experimental work and provides discussion of results, and section five provides the conclusions and future works directions.

## 2 Literature Review

Medical imaging has been rapidly innovated in the last decades, and the resolution of scanners reflect a huge amount of acquired data. Therefore, huge image data is available in electronic health systems. As for the important role that medical images played in diagnosis applications, the diagnosis process should be executed more accurate, in real time, and based on 3D medical volumes acquired from medical scanners.

It has been defined that Objects of Interest (OOI) are presented by the occupied volume or by surrounding edges. But in radiologist field, diagnosing 3D medical images focuses on classifying voxels into parts with similar properties. In 2010, We introduced a system that does 3D medical volume segmentation using hybrid multi-resolution statistical approaches. In 2011, we conducted an experiment on multi-resolution analysis using wavelet, ridgelet, and curvelet transforms for medical image segmentation. Then, we applied the proposed segmentation techniques on a reduced feature medical data us- ing PCA [8, 52, 78, 44]. We enhance the application of HMMs for accelerating medical volumes segmentation in [22, 4, 31, 62] and compare the achieved results to different MRA techniques. The problem associated with the best segmentation method is the processing time complexity. In 2011, we introduced a parallel processing solution to address this problem. In 2016, we executed the segmentation process on GPU to detect breast cancer using single-pass Fuzzy C-Means clustering algorithm.

In 2017, we proposed a robust and accurate intelligent system to enhance 3D segmentation techniques for reconstructed 3D medical volumes, and accelerated the 3D medical volume segmentation using GPUs. The implemented algorithm has been improved efficiently and a more accurate and faster algorithm was introduced in [43, 9, 30, 58]. In 2018, we proposed a multi-orientation geometric medical volumes segmentation using 3d multi-resolution analysis, where multiple orientation views were experimented to achieve the most accurate results [42, 46, 50, 65]. A new dataset was used in the experiment to validate the segmentation process. Execution time complexity was resolved using hardware accelerators (GPU). In [32, 16, 73, 55], we introduce Transfer Learning to accelerate the segmentation process. We propose a methodology for transferring HMM matrices from image to another skipping the training time for the rest of the 3D volume. One HMM train is generated and generalized to the whole volume.

Many other research groups worked previously in the volumetric medical imaging such in [6, 47, 56]. Some of them have worked in multimedia networking and medical image segmentation acceleration. Ali et al propose in [60, 35, 38, 37] a novel routing protocol for multimedia sensor protocols. This protocol was based on the selection of the next hop node that has the highest throughput and the node that is closer to the destination node using a greedy method. The protocol proved its superiority when compared to its competitor protocols and showed its efficiency for large scale multimedia data. Gek Hong et al propose in [81, 72, 77, 51] an improvement to the video transmission of high data rates over a wireless multimedia sensor network. For this sake, they used a bit-stream partitioning algorithm for unequal error protection (UEP) of video signals. Furthermore, adaptive equal-unequal error protection was also proposed. Wei et al introduce in [70, 55, 6, 47] Multimedia Information Networks (MINets), and they incorporate visual and textual information in their work. Their goal was to use deep learning algorithms for realistic multimedia applications.

An efficient CNN framework for medial image segmentation had been provided by Xiuxia et al [89, 36, 15, 39]. This framework avoided the long time execution and being fault vulnerable. As part of their work, they use a distributed cross-platform parallelism implementation. Moreover, they also use a Theano's GPU implementation along with recommending the use of multi-core CPUs and many-core Intel Phi. Patrick Nigri et al present in [54, 45, 57, 12] an efficient segmentation method that is capable of handling very large high resolution images. The solution was based on the MapReduce model, and it was implemented and validated using the Hadoop platform. The experimental work proved the efficiency and the scalability of the model. Hanchuan et al provide in [76, 79, 11, 67] a new hierarchical distributed genetic algorithm for image segmentation. This method does not require a prior number of image regions. Furthermore, a more compact format was used to ensure a better storage efficient implementation. Finally, the fitness function was revised as well.

Recently, deep learning methods have opened the door wide open to improve medical image segmentation. Examples works include Zhang et al, who proposed a method for domain generalization using deep transformation. They used the deep learning on huge augmented data, and later, this algorithm can generalize well on unseen domains [88, 59, 2, 19]. Sun et al provided an efficient method for 3D segmentation of pulmonary nodules based using semi-supervised learning. As part of their work, three parallel convolutions were utilized [83, 17, 20]. Sleman et al Introduced a 3D segmentation method based on an appearance model for 3D OCT data and an adaptive patient-specific retinal atlas [82, 13, 14]. Martin et al proposed a 3D reconstruction method to estimate cerebral ventricle system (CVS) using deep learning. They argued that the estimation using 2D methods is error-prone. They evaluated the segmentation accuracy using Dice, Hausdorff distance (dH) [71]. Guo et al proposed an approach for 3D tumor segmentation using deep learning. Tumor detection is an important process in tumor analysis. In details, Deep LOGIS-MOS approach for 3D tumor segmentation was proposed. As part of their work, a deep contextual learning for boundaries were used. The deep learning structure is composed of a fully convolutional network [53].

## 3 Methodology

Segmenting 3D medical volumes using HMM over a secure distributed multimedia network has been proposed in this research paper. The following sections describe how our proposed system could be applied to acquire an accurate segmented volumes in real-time.

### 3.1 Proposed System

The 2D plane medical images are extracted from the acquisition systems such as (PET, CT, MRI). It is common for the patient to lie in the axial orientation.

The result would be many 2D images depending on the thickness of the bed movements and the scanned part of the patient body. These images are used to build the 3D volume which will be used in the proposed work.

The main idea of the proposed system is to divide the workload into parts that can be executed over distributed devices securely, the segmented parts will be gathered back to the server. In terms, it illustrates the segmented volume to the end-user. Figure 1 illustrates the proposed system and its components.

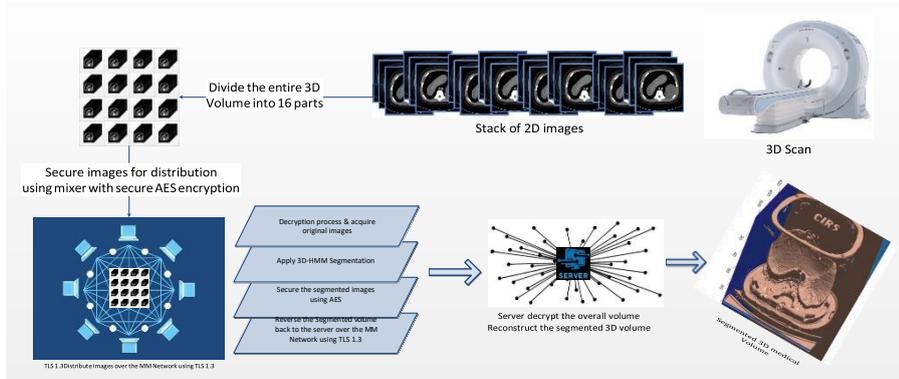

**Fig. 1** The Proposed System

From Figure 1, it can be noticed that extra workload has been added to the overall execution for applying the segmentation process. But the overall execution time should be less than many other HMM procedures of segmentation processes.

### 3.2 Statistical Segmentation - HMM

HMM has proved its efficiency in medical image segmentation. However, it suffers from poor execution time. According to [40]; HMMs is one of the ideal medical image segmentation techniques compared to the work done in other methods such as k-means and traditional thresholding. However, the problem still significant according to the required computational time for training the HMM to generate the probability matrices.

The weakness of HMMs is its long processing time, compared with the other available segmentation techniques. AlZu'bi et al experimented NEMA IEC body phantom to validate the segmentation accuracy using HMMs and MRA for medical image segmentation [61]. The long computational time of training HMMs was raised here, where multiple doubles of the required time for applying other supervised segmentation techniques needed to segment medical data using HMMs.

Many researchers have worked on such issues for acceleration [10, 74, 91, 90, 64, 63]. GPU-based parallel processing has been used as an efficient hard-ware solution, it is still inadequate for 3D volumes due to the huge amount     of data in 3D volume and the large stack of Medical DICOM files that pose     an overhead on the training process of HMMs. AlZu'bi et al accelerated the HMM segmentation process in [32] using the transfer learning. The Algorithm explained in Figure 2 explains the procedure for segmenting 2D images us-    ing HMMs [27] and the blocks which slow down the segmentation time are highlighted.

```
Read the Image
Get the size of the image
initiate a counter to 0
Define the sequence of feature vectors      -    seq(counter)
 Apply HMM viterbi
              likelystates = hmmviterbi(seq, TRANS_EST2, EMIS_EST2);
Estimate New Matrices
              [TRANS_GUESS,EMIS_GUESS] = hmmestimate(seq,likelystates);
Train to obtain more accurate Matrices
              [TRANS_EST2, EMIS_EST2] = hmmtrain(seq, TRANS_GUESS, EMIS_GUESS);
Iterate Viterbi for most likely states
              likelystates = hmmviterbi(seq, TRANS_EST2, EMIS_EST2);

Segment the image Based on the likelystates
Show the Segmented image
```

**Fig. 2** The procedure for segmenting 2D images using HMMs with highlighted slow parts

## 3.3 Secure Multimedia Distributed Network

The proposed system aims at applying the code presented in Figure 2 over distributed devices. Acquiring segmented volume from one device is deemed slow with respect to the training steps of HMM. We divide the entire patient volume into a set of sub-volumes that have been distributed over the network into trusted devices to do concurrent segmentation using HMMs.

The methodology of securing image distribution over the multimedia network has been done through the following steps:

- **Splitting the original image into 16 parts (based on available devices).** In this stage, we locate the original image in a specific location to be loaded through the splitter program. Then, read the image and split     it using a specific matrix that contains the required rows and columns. We Determine a (4*4) array with to hold the image chunks, and initialize the image array with the available chunks. Finally, we draw the image chunks using Graphic class and reorder these random chunks into the exact order   to complete the splitting  stage.
- **Encrypt each part of the original image using the secure AES encryption algorithm.** Starting with getting the user's password. Arbi-

trary salt data, has been used to make guessing attacks against the pass-word more difficult to pull off. Then, get the image parts one by one from the specific direction and apply a PBE cipher object and initialize for en-cryption. Finally, start ciphering the image parts using the specified AES encryption method and complete the ciphering stage.

- **Start distributing the encrypted image parts over the multimedia network based on predeftned order to the other authenticated end in a secure manor.** Transport Layer Security (TLS) is the most widely-deployed security protocol, it provides authentication, privacy, and data integrity. TLS 1.3 is employed here to reduce the chance of implementation errors, and remove features no longer needed.
- **Decrypt the image parts using the same encryption technique by the authenticated user.** All distributed trusted devices decrypt the transmitted images and form a PBE cipher object and initialize it for de-cryption. the trusted devices read the image using the cipher input stream, which wraps a file input stream. Then, write the decrypted image out using AES decryption mode.
- **Reorder the image parts into the agreed order and combine them to have the original image.** At this stage, the distributed devices can start the statistical segmentation process using HMM concurrently to save execution time.
- **Reversing the segmented parts securely.** The segmented results will be encrypted using the same encryption technique and redirected backward to the server.
- **Segmented volume reconstruction.** The server will decrypt the over-all volume which is partially segmented by distributed trusted devices and reconstruct the segmented 3D volume. The reconstructed volume is seg-mented and OOI should be illustrated clearly as an isolated volume.

## 4 Results and Analysis

The goal of our proposed system is to enhance the quality of diagnosis in im-ages, especially in segmenting 3D medical volumes. The following subsections prove the superiority of this method in comparison with other systems.

### 4.1 Testing Data (CIRS abdominal phantom, model 057 [49])

The abdominal phantom model 057 was used to generate the phantom exper-iments [49] (shown in Fig. 10). This model was produced from Computerized Imaging Reference Systems (CIRS). Based on CIRS group in [49], a formula-tion of Zerdine ℝ, used for self-healing, has been utilized to create the CIRS Triple Modality Phantom. This contributed to making the process of inserting a small needle easier, and importantly improve the navigation demonstration of images [49].

This phantom plays an important role in diagnosing and monitoring of patient treatments. This explains the use of the presented phantom in medical imaging developments, scanning methodologies, testing, validating and the demonstration of medical imaging application systems [87]. This model facilitates simulating a small adult abdomen [48]. In consequence, this phantom will be adopted in this project to validate the proposed techniques. The model 057A uses simplified anthropomorphic geometry to provide an artificial simulation of the area from the thorax vertebrae (T9/T10) to the lumbar vertebrae (L2/L3) in the abdomen. The materials are made to provide contrast between the structures under scanners circumstances to prevent the background gel from leakage during the scanning process.

The experimented phantom includes simulated parts, such as liver, part of the lung, abdominal aorta, the vena cava, some ribs, and a spine. There are ten different obstacles representing the lesions spreading in this phantom, six of which are located in the liver, two identical lesions in kidneys (one in each), and two different lesions in the muscle layer and outside fat layer which surround the simulated parts. The abdominal phantom model 057 is made up by plastic end caps to be durable for multiple scans [84, 49, 87]. The inserted simulated lesions are with high contrast to the model background [49]. These features increase the performance of evaluating the targeting accuracy of the proposed system.

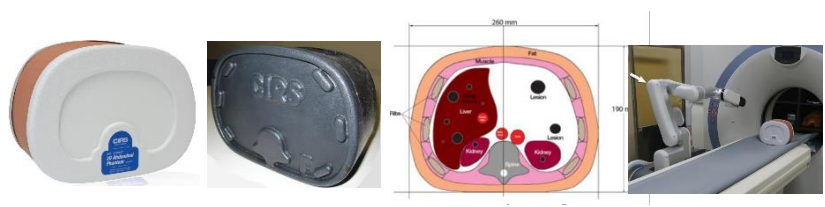

**Fig. 3** CIRS abdominal phantom, model 057 [84, 49, 87]

Fiducial markers were used to collect the preoperative CT dataset of this abdomen phantom [48]. These markers were adopted as they provide a strong point-based registration [66]. The preregistered images were obtained using CT scanner [66]. This dataset was used by AlZubi et al. in [43] in order to evaluate their 3D segmentation system. A comparison was made by J. von Berg et al. in [48] between reformatted CT slice and ultrasound image in realtime by overlaying both kinds of information directly (Fig. 11) [66] . It is worth noting that some preprocessing steps such as image denoising were used to visualize the result properly. In order to compare the manual and auto- matic registration approaches, the root mean square of all Cartesian distances between corresponding positions was considered. Furthermore, a regular grid of 277 reference positions distributed over the plane (the ultrasound fan) was used in both images [48].

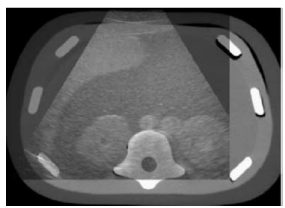

**Fig. 4** Real-time ultrasound image overlaid with corresponding multiplanar reformatted CT slice for the CIRS abdominal phantom (model-057) [48]

## 4.2 Security

The protection mechanism used here added a new security level over the transition process and the data at rest. By providing a robust encryption algorithm with a splitting algorithm to each image of the patient. Implementing the proposed system over multimedia networks leave both theoretical and practical impacts. It can theoretically contributed to improve the 3D image segmentation over those networks. Furthermore, Image transportation along multimedia networks are risky, especially with medical data, and requires a highly demanded environment. This suffers from the lack of works and the limitation of the performance. Practically, this work would significantly help the medical sector by improving the accuracy of medical diagnosis.

The Advance Encryption Standard (AES) method was employed in this research to encrypt the image chunks. The AES put the image data into an array; after which the cipher transformations are repeated over several encryption rounds depending on the key length. The AES has proven to be unthreatened due to its high-computational complexity. The early testing stage in this research was made using a 4*4 matrix that splits the images into 16 parts. Other matrices could be used here, but with affecting the results ac- cording to execution time latency, accuracy, and algorithm complexity. Table 1 shows how the matrix size affect the encryption process.

**Table 1** Security evaluation measurements

| Matrix size | Algorithm complexity | Required encryption time (ms) | Encryption accuracy |
|---|---|---|---|
| 1 X 1 (Original) | 20% | 0.002 | 55% |
| 3 X 3 | 30% | 0.030 | 60% |
| 5 X 5 | 50% | 0.080 | 65% |
| 7 X 7 | 70% | 0.140 | 70% |
| 9 X 9 | 90% | 0.330 | 80% |

From Table 1, measurement values are indicating a percentage scale according to algorithm complexity and encryption accuracy. In algorithm complexity, 20% is the least complexity which is by encrypting the original medical image without any splitting. 90% is the most complex encryption criteria which has been achieved using the largest splitter of 81 parts. The required encryption time is measured in milliseconds before any secure transmission. It is worth mentioning here that while the complexity of the encryption technique is increased, the execution time should also be increased in the same vein. It can be noticed from Table 1 that the encryption accuracy is getting better while the matrix size is bigger.

Different encryption procedures have been used to secure image transmission over distributed multimedia networks. As mentioned in Table 1, each procedure affects the proposed system validity. Figure 4.2 illustrates how we encrypt the transmission.

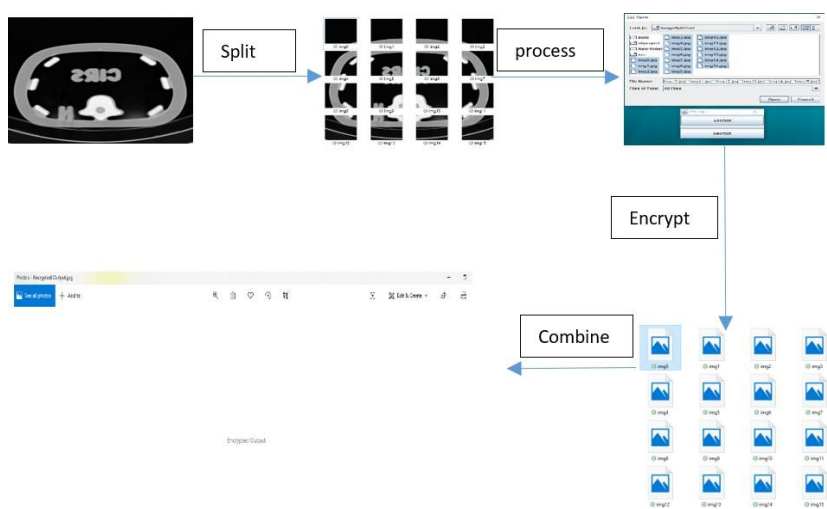

**Fig. 5** Encrypting medical images using the proposed security system

### 4.3 Accuracy measurements

To validate the proposed system, we use two validation ways and compare the results we gained in this work to results we achieved in our previous work once [33], and to other researchers results on the same dataset [49].

The following is a list of accuracy measurements used in this work, and details of each.

– **Real patient data with mysterious lesions.** The achieved results after applying the proposed system on real patient images have been reviewed by radiologists. Their opinions were significant to assess the proposed system, they answered a questionnaire based on the achieved segmentation results.

**Table 2** 19 different radiologists point of view for applying our proposed system on real datasets

| Questions answered by radiologists | (1) Strongly Disagree | (2) Disagree | (3) Maybe | (4) Agree | (5) Strongly Agree |
|---|---|---|---|---|---|
| Abnormalities are defined | 0 | 0 | 4 | 8 | 10 |
| Wide range of features included | 3 | 2 | 11 | 6 | 0 |
| Is it easily to miss ROI by radiologists | 22 | 0 | 0 | 0 | 0 |
| Is it easily to misinterpret ROI by radiologists | 21 | 1 | 0 | 0 | 0 |
| Is reading large number of mammographic images becomes harder | 14 | 3 | 3 | 1 | 1 |
| Does the proposed method better than screening programs | 2 | 6 | 3 | 8 | 3 |
| Does the proposed method help in provide an accurate diagnosis | 0 | 0 | 6 | 15 | 1 |
| Does the proposed method help reducing the number of false positives | 0 | 0 | 7 | 15 | 0 |
| Does the proposed method assist in deciding between follow up and biopsy | 1 | 3 | 7 | 11 | 0 |
| Can the proposed system replace the nowadays available systems | 0 | 3 | 17 | 2 | 0 |

Table 2 illustrates the statistics of 22 different radiologists assessment of our proposed system.

From Table 2, the proposed system can be accepted with all radiologists. 96% of samples seen the system as easy and beneficial in their daily work. Over 63% of them overcame our system over the traditional. The achieved accuracy of the diagnosis process has been increased according to 86% of the sample. Finally, 84% of the sampled radiologists agreed that our proposed system can replace the nowadays available systems.

– **Simulated phantom data.** CIRS phantom [49] which has been illustrated in Figure 3, and NEMA IEC phantom [40, 61] have been used to validate the proposed system. All inserted objects to the phantom have been segmented, and measured in many segmentation techniques includ- ing the proposed one.

Table 3 illustrates the achieved results after applying the proposed techniques on NEMA IEC phantom, and compare them to other previous results. Euclidean Distance (ED) has been used to measure the spheres diameters and calculate errors according to equation 1.

$$[h]error\ \% = \frac{MeasuredDiameter - ActualParameter}{ActualParameter} \times 100\% \quad (1)$$

From Table 3, promising results have been achieved on detecting NEMA IEC phantom spheres using the proposed techniques. In fact, the three in- troduced methodologies in [42] produced the best achieved results, hence, we are mostly using the same technique. The complexity of the execution time is not considered in this table, which is the strength of our proposed system to overcome the previous work proposed in [42]. NEMA IEC phantom alone is not satisfying alone for validating the proposed technique. Therefore, CIRS Model [49] (illustrated in Figure 3) is employed here to validate the proposed system. Table 4 presents the achieved results of the inserted spheres into CIRS model 057 phantom after the segmentation process using the proposed system, and compare the achieved results to what we achieved in [42]. Table 4 shows the validity of segmenting CIRS phantom using our proposed technique and the previously implemented system in [42]. Before getting the results from Table 4, we identified the relation between the spheres group in the phantom. The CIRS Model 057 phantom contains 4 groups of inserted spheres as explained in Figure 6. These analytical data can help in accurately validate the proposed techniques. Obviously, Table 4 proves the validity and efficiency of the proposed system in term of segmentation accuracy. The segmentation accuracy of using HMM on 8 Distributed Devices in a multimedia network, with (4x4) AES matrix splitting encryption is 99.001. Whereas, the segmentation accuracy of using HMM on 8 Distributed Devices in a multimedia network, with (9x9) AES matrix splitting encryption is 98.685%. The segmentation ac-curacy of using HMM on 16 Distributed Devices in a multimedia network, with (4x4) AES matrix splitting encryption is 98.994%. Whilst, the segmentation accuracy of using HMM on 16 Distributed Devices in a multimedia network, with (9x9) AES matrix splitting encryption is 98.544%. In details, the average accuracy of the proposed system for detecting the overall spheres in CIRS phantom is 98.794, which is sufficient accuracy in such segmentation applications. The achieved accuracy can be easily improved if we ignore the achieved results using AES encryption with a (9x9) matrix for the split. This matrix splitting effect negatively on the achieved result and increased the segmentation error. Nevertheless, our proposed system overcomes the other methods by execution time complexity. From Table 4, it can be noticed that the proposed system accuracy is decreasing at the AES splitting using a (9x9) matrix. This is due to the added noise in the partitioning and rearrangement steps.

**Table 3** The error percentages of spheres measurements using different segmentation techniques for NEMA IEC body phantom [40, 61]

| Spheres (mm) | | Error % for measured diameters | | | | | |
|---|---|---|---|---|---|---|---|
| | | 10 mm | 13 mm | 17 mm | 22 mm | 28 mm | 37 mm |
| K-means [29] | | 13.6 | 11.5 | 5.77 | 5.51 | 5.1 | 5.01 |
| MRFM [29] | | 7.41 | 8.69 | 4.28 | 4.06 | 3.9 | 3.89 |
| Clustering [29] | | 18.6 | 16.0 | 9.0 | 7.5 | 5.5 | 1.1 |
| Iterative Thresholding [29] | | 3.0 | 3.1 | 0.6 | 0.9 | 1.1 | 1.8 |
| Haar Wavelet filter [23] | Level 1 (Axial) | -2.9 | -2.46 | 1.35 | 0.82 | 0.29 | 0.05 |
| | Level 2 (Axial) | -10.9 | -6.67 | 3.88 | -1.3 | -0.76 | -1.95 |
| | Level 3 (Axial) | — | — | 5.65 | -18.2 | 2.57 | -3.24 |
| 3D Geometric Discrete Wavelet Transform (DWT) [42] | Axial | 1.85 | 1.53 | 1.51 | 0.69 | 0.17 | 0.05 |
| | Sagittal | 1.92 | 1.47 | 1.44 | 0.58 | 0.22 | 0.06 |
| | Coronal | 1.77 | 1.39 | 1.25 | 1.01 | 0.19 | 0.06 |
| | Multi-Orientation | 1.81 | 1.40 | 1.31 | 0.65 | 0.19 | 0.05 |
| Proposed System | HMM 4*4-splitted AES 8 distributed devices | 1.72 | 1.41 | 1.49 | 0.65 | 0.14 | 0.05 |
| | HMM 9*9-splitted AES 8 distributed devices | 2.01 | 1.80 | 1.72 | 0.93 | 0.28 | 0.08 |
| | HMM 4*4-splitted AES 16 distributed devices | 1.68 | 1.38 | 1.44 | 0.63 | 0.12 | 0.05 |
| | HMM 9*9-splitted AES 16 distributed devices | 2.03 | 1.82 | 1.71 | 0.95 | 0.15 | 0.05 |

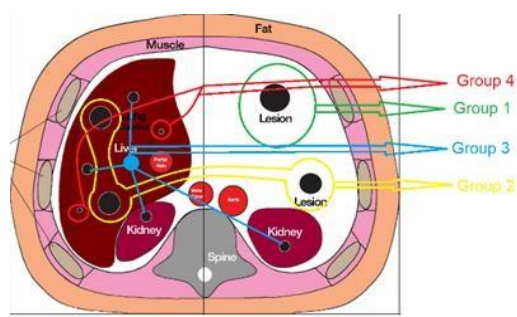

**Fig. 6** CIRS Model 057 phantom - inserted lesion groups

**Table 4** The error percentages of lesions measurements using the proposed segmentation technique for CIRS Model 057 Phantom [49]

| Spheres | | Error % for measured Lesions | | | | | | | | | |
|---|---|---|---|---|---|---|---|---|---|---|---|
| | | Grp 1 | Grp 2 | | | Grp 3 | | | | Grp 4 | |
| | | Sph1 | Sph1 | Sph2 | Sph3 | Sph1 | Sph2 | Sph3 | Sph4 | Sph1 | Sph2 |
| **3D Geometric DWT** | Axial | 0.03 | 0.07 | 0.06 | 0.08 | 1.01 | 1.03 | 1.10 | 1.01 | 3.50 | 3.85 |
| | Sagittal | 0.04 | 0.07 | 0.08 | 0.08 | 1.03 | 1.08 | 1.11 | 0.99 | 3.65 | 4.05 |
| | Coronal | 0.03 | 0.06 | 0.08 | 0.07 | 1.07 | 1.05 | 1.09 | 1.02 | 3.05 | 3.75 |
| | Multi Orientation | 0.03 | 0.06 | 0.06 | 0.07 | 1.02 | 1.04 | 1.10 | 1.00 | 3.45 | 3.90 |
| **3D Geometric HMMs** | Axial | 0.03 | 0.06 | 0.05 | 0.06 | 0.93 | 1.00 | 1.01 | 0.96 | 2.74 | 3.05 |
| | Sagittal | 0.03 | 0.07 | 0.06 | 0.06 | 0.94 | 1.01 | 1.07 | 0.99 | 2.85 | 3.11 |
| | Coronal | 0.03 | 0.05 | 0.06 | 0.05 | 0.91 | 0.98 | 1.05 | 1.02 | 2.93 | 2.98 |
| | Multi Orientation | 0.03 | 0.06 | 0.05 | 0.05 | 0.92 | 0.99 | 1.06 | 0.97 | 2.81 | 3.01 |
| *Proposed Model* | | | | | | | | | | | |
| **HMM 8 Distributed Devices** | AES (4*4) matrix split | 0.03 | 0.06 | 0.06 | 0.07 | 0.95 | 1.01 | 1.03 | 0.98 | 2.82 | 2.98 |
| | AES (9*9) matrix split | 0.04 | 0.09 | 0.08 | 1.00 | 0.99 | 1.12 | 1.31 | 1.22 | 3.80 | 3.50 |
| **HMM 16 Distributed Devices** | AES (4*4) matrix split | 0.03 | 0.06 | 0.05 | 0.07 | 0.92 | 0.99 | 1.06 | 0.97 | 2.88 | 3.03 |
| | AES (9*9) matrix split | 0.04 | 0.10 | 0.08 | 1.00 | 1.10 | 1.24 | 1.44 | 1.65 | 3.92 | 3.99 |

## 4.4 Discussion

According to the experimental work and results, the proposed system proved to be comparable to other state-of-the-art segmentation methods in term of accuracy. Specifically, its accuracy of 98.994% was comparable to that of 3D Geometric DWT. However, our proposed system has two advantages over that method and other top performers. First, the use of AES encryption has provided a secure environment, which would be critical in many domains including the medical one, as these medical images are considered private data. The use of (4x4) window provided a trade off between the encryption accuracy and the encryption time. Clearly, the selection of the window size is domain-dependent, as the importance of time and the accuracy varies across different domains. Second, the use of distributed environment reduced the segmentation time significantly, as it applies the concept of distributed processing and it divides the load into various tasks that can be executed concurrently.

It was noted that having smaller window size, although would incur a higher encryption error, but would result in a better segmentation accuracy. This can be easily seen in Table 3 and Table 4. This is reasonable as smaller window sizes are more capable of capturing smaller volume details. Furthermore, it was noted that the use of more distributed devices, although would provide a faster execution, would slightly decrease the segmentation accuracy. In addition, it was noted that the more the diameter of the object, the more it is easier to be detected, and thus, the better the segmentation accuracy. This can be due to the fact that larger objects are less sensitive to noise.

It is worth mentioning that many parts of our proposed system can be further optimized. First, many other encryption methods can be compared to find the optimal one. Second, many alternative segmentation methods can be used along with HMM to further optimize the segmentation accuracy. Third, some other preprocessing techniques can be used to provide a fast execution in certain scenarios. These optimization strategies can provide promising results and are left as part of the future work. A prior knowledge of the nature of the data would be contribute significantly in selecting the appropriate thresholds, which in turn would provide a better accuracy. This work would have both theoretical and practical impacts on multimedia networks. Theoretically, it would contribute in further improving the research on segmenting 3D images, which suffers from the lack of works and the limitation of the performance.

Practically, this work would significantly help the medical sector by improving the accuracy of medical diagnosis.

## 5 Conclusion and future work

In this work, an efficiently secure 3D image segmentation method was proposed. The method uses HMM segmentation method in a distributed environment in order to overcome the overhead of 3D segmentation and improve the execution time of the method. Moreover, it uses AES encryption method to secure the data in such distributed environment. From the experimental work, it was proved that the use of HMM in a distributed environment has achieved accuracy up to 98.994%, which is comparable to the state of art segmentation methods, while our proposed method outperformed the existing methods in term of security and execution time. The use of AES with (4x4) window has obtained an encryption accuracy of 62% with a negligible encryption time of 0.06ms. As for the execution time, it is obvious that the load distribution would contribute significantly in reducing the execution time of 3D image segmentation.

Future work can be conducted in many directions. First, more encryption methods can be used for comparison purposes to further improve the security level. Second, other segmentation methods can also be used, along with HMM, to optimize the segmentation accuracy. Third, the distributed environment can be optimized using software and hardware solutions to provide optimal segmentation performance.

## Abbreviations

| | |
|---|---|
| **2D** | 2 Dimensional |
| **3D** | 3 Dimensional |
| **MIP** | Medical Image Processing |
| **HMM** | Hidden Markov Models |
| **CAD** | Computer Added Diagnosis |
| **AES** | Advanced Encryption Standard |
| **OOI** | Object Of Interest |
| **ROI** | Region Of Interest |
| **GPU** | Graphical Processing Unit |
| **CPU** | Central Processing Unit |
| **MINets** | Multimedia Information Networks |
| **CNN** | Convolutional Neural Network |
| **PET** | Positron Emission Tomography |
| **CT** | Computed Tomography |
| **MRI** | Magnetic Resonance Imaging |
| **IEC** | The International Electrotechnical Commission |
| **NEMA** | National Electrical Manufacturer Association |
| **MRA** | Multi-Resolution Analysis |
| **DICOM** | Digital Imaging and COmmunications |
| **TLS** | Transport Layer Security |
| **CIRS** | Computerized Imaging Reference Systems |

## References

1. Abooraig R, Al-Zu'bi S, Kanan T, Hawashin B, Al Ayoub M, Hmeidi I (2018) Automatic categorization of arabic articles based on their political orientation. Digital Investigation 25:24 – 41

2. Abusukhon A, AlZu'bi S (2020) New direction of cryptography: A re- view on text-to-image encryption algorithms based on rgb color value. In: 2020 Seventh International Conference on Software Defined Systems (SDS), IEEE, pp 235 – 239

3. Adamo F, Affanni A, Agethen R, Al Osman H, Alecci A, Alemanno A, Aliahmad B, AlZubi S, Ancona A, Andria G, et al (????) Abbafati, manuel 376 abbod, maysam 188, 619 abousharkh, maha 257 abundo, paolo 593

4. Al-Ayyoub M, AlZu'bi SM, Jararweh Y, Alsmirat MA (2016) A gpu-based breast cancer detection system using single pass fuzzy c-means clustering algorithm. In: 2016 5th International Conference on Multimedia Comput- ing and Systems (ICMCS), IEEE, pp 650 – 654

5. Al-Ayyoub M, AlZu'bi S, Jararweh Y, Shehab MA, Gupta BB (2018) Ac- celerating 3d medical volume segmentation using gpus. Multimedia Tools and

Applications 77(4):4939 – 4958

6. Al-Zoubi H, Al-Zu'bi S, Stamatakis S, Almimi H (2018) Ruled surfaces of finite chen-type. Journal for Geometry and Graphics 22(1):015 – 020

7. Al Zu'bi S, Islam N, Abbod M (2010) 3d multiresolution analysis for re- duced features segmentation of medical volumes using pca. In: Circuits and Systems (APCCAS), 2010 IEEE Asia Pacific Conference on, IEEE, pp 604 – 607

8. Al-Zu'bi S, Al-Ayyoub M, Jararweh Y, Shehab MA (2017) Enhanced 3d segmentation techniques for reconstructed 3d medical volumes: Robust and accurate intelligent system. Procedia computer science 113:531 – 538

9. Al-Zu'bi S, Al-Ayyoub M, Jararweh Y, Shehab MA (2017) Enhanced 3d segmentation techniques for reconstructed 3d medical volumes: Robust and accurate intelligent system. Procedia Computer Science 113:531 – 538, DOI https://doi.org/10.1016/j.procs.2017.08.318, URL http://www.sciencedirect.com/science/article/pii/S1877050917317283, the 8th International Conference on Emerging Ubiquitous Systems and Pervasive Networks (EUSPN 2017)

10. Al-Zu'bi S, Hawashin B, Mughaid A, Baker T (2020) Efficient 3d med- ical image segmentation algorithm over a secured multimedia network. Multimedia Tools and Applications pp 1 – 19

11. Alasal SA, Alsmirat M, Baker QB, Alzu'bi S, et al (2020) Lumbar disk 3d modeling from limited number of mri axial slices. International Journal of Electrical and Computer Engineering 10(4):4101

12. AlKhatib AA, Sawalha T, AlZu'bi S (2020) Load balancing techniques in software-defined cloud computing: an overview. In: 2020 Seventh International Conference on Software Defined Systems (SDS), IEEE, pp 240–244

13. Alsmadi A, AlZu'bi S, Al-Ayyoub M, Jararweh Y (2020) Predicting helpfulness of online reviews. arXiv preprint arXiv:200810129

14. Alsmadi A, AlZu'bi S, Hawashin B, Al-Ayyoub M, Jararweh Y (2020) Employing deep learning methods for predicting helpful reviews. In: 2020 11th International Conference on Information and Communication Sys- tems (ICICS), IEEE, pp 007–012

15. AlZubi S (2011) 3d multiresolution statistical approaches for accelerated medical image and volume segmentation. PhD thesis, Brunel University School of Engineering and Design PhD Theses

16. AlZubi S, Amira A (2010) 3d medical volume segmentation using hybrid multiresolution statistical approaches

17. AlZu'bi S, Jararweh Y (2020) Data fusion in autonomous vehicles research, literature tracing from imaginary idea to smart surrounding community. In: 2020 Fifth International Conference on Fog and Mobile Edge Comput- ing (FMEC), IEEE, pp 306–311

18. AlZu'bi S, AlQatawneh S, ElBes M, Alsmirat M (????) Transferable hmm probability matrices in multi-orientation geometric medical volumes segmentation. Concurrency and Computation: Practice and Experience p e5214

19. AlZubi S, Islam N, Abbod M (2011) Enhanced hidden markov models for accelerating medical volumes segmentation. In: GCC Conference and Exhibition (GCC), 2011 IEEE, IEEE, pp 287–290

20. AlZubi S, Islam N, Abbod M (2011) Multiresolution analysis using wavelet, ridgelet, and curvelet transforms for medical image segmentation. International journal of biomedical imaging 2011

21. AlZubi S, Sharif MS, Abbod M (2011) Efficient implementation and evaluation of wavelet packet for 3d medical image segmentation. In: 2011 IEEE international symposium on medical measurements and applications, IEEE, pp 619–622

22. AlZubi S, Sharif MS, Islam N, Abbod M (2011) Multi-resolution analysis using curvelet and wavelet transforms for medical imaging. In: 2011

IEEE international symposium on medical measurements and applications, IEEE, pp 188 – 191

23. AlZubi S, Jararweh Y, Shatnawi R (2012) Medical volume segmentation using 3d multiresolution analysis. In: 2012 international conference on innovations in information technology (IIT), pp 156 – 159

24. AlZu'bi S, Shehab MA, Al-Ayyoub M, Benkhelifa E, Jararweh Y (2016) Parallel implementation of fcm-based volume segmentation of 3d images. In: 2016 IEEE/ACS 13th International Conference of Computer Systems and Applications (AICCSA), IEEE, pp 1 – 6

25. AlZu'bi S, Al-Qatawneh S, Alsmirat M (2018) Transferable hmm trained matrices for accelerating statistical segmentation time. In: 2018 Fifth International Conference on Social Networks Analysis, Management and Security (SNAMS), IEEE, pp 172 – 176

26. AlZu'bi S, Hawashin B, ElBes M, Al-Ayyoub M (2018) A novel recommender system based on apriori algorithm for requirements engineering. In: 2018 fifth international conference on social networks analysis, management and security (snams), IEEE, pp 323 – 327

27. AlZu'bi S, Alsmadiv A, AlQatawneh S, Al-Ayyoub M, Hawashin B, Jararweh Y (2019) A brief analysis of amazon online reviews. In: 2019 Sixth International Conference on Social Networks Analysis, Management and Security (SNAMS), IEEE, pp 555 – 560

28. AlZu'bi S, Alsmirat M, Al-Ayyoub M, Jararweh Y (2019) Artificial intelligence enabling water desalination sustainability optimization. In: 2019 7th International Renewable and Sustainable Energy Conference (IRSEC), IEEE, pp 1 – 4

29. AlZu'bi S, Aqel D, Mughaid A, Jararweh Y (2019) A multi-levels geo-location based crawling method for social media platforms. In: 2019 Sixth International Conference on Social Networks Analysis, Management and Security (SNAMS), IEEE, pp 494 – 498

30. Alzu'bi S, Badarneh O, Hawashin B, Al-Ayyoub M, Alhindawi N, Jararweh Y (2019) Multi-label emotion classification for arabic tweets. In: 2019

Sixth International Conference on Social Networks Analysis, Management and Security (SNAMS), IEEE, pp 499 – 504

31. Alzubi S, Hawashin B, Mughaid A, Jararweh Y (2020) Whats trending? an efficient trending research topics extractor and recommender. In: 2020 11th International Conference on Information and Communication Systems (ICICS), IEEE, pp 191 – 196

32. AlZu'bi S, Jararweh Y, Al-Zoubi H, Elbes M, Kanan T, Gupta B (2018) Multi-orientation geometric medical volumes segmentation using 3d mul- tiresolution analysis. Multimedia Tools and Applications pp 1 – 26

33. AlZu'bi S, Shehab M, Al-Ayyoub M, Jararweh Y, Gupta B (2018) Parallel implementation for 3d medical volume fuzzy segmentation. Pattern Recognition Letters DOI https://doi.org/10.1016/j.patrec.2018.07.026, URL http://www.sciencedirect.com/science/article/pii/S016786551830326X

34. AlZu'bi S, Hawashin B, Mujahed M, Jararweh Y, Gupta BB (2019) An efficient employment of internet of multimedia things in smart and future agriculture. Multimedia Tools and Applications 78(20):29581 – 29605

35. AlZu'bi S, Mughaid A, Hawashin B, Elbes M, Kanan T, Alrawashdeh T, Aqel D (2019) Reconstructing big data acquired from radioisotope distribution in medical scanner detectors. In: 2019 IEEE Jordan International Joint Conference on Electrical Engineering and Information Technology (JEEIT), IEEE, pp 325 – 329

36. Aqel D, AlZu'bi S, Hamadah S (2019) Comparative study for recent technologies in arabic language parsing. In: 2019 Sixth International Confer- ence on Software Defined Systems (SDS), IEEE, pp 209 – 212

37. Berg JV, Kruecker J, Schulz H, Meetz K, Sabczynski J (2004) A hybrid method for registration of interventional ct and ultra-sound images. International Congress Series 1268:492 – 497, DOI https://doi.org/10.1016/j.ics.2004.03.171

38. Computerized Imaging Reference Systems I (2013) Triple modality 3D abdominal phantom, Model 057A. CIRS

39. Elbes M, Almaita E, Alrawashdeh T, Kanan T, AlZu'bi S, Hawashin B (2019) An indoor localization approach based on deep learning for indoor location-based services. In: 2019 IEEE Jordan International Joint Conference on Electrical Engineering and Information Technology (JEEIT), IEEE, pp 437 – 441

40. Elbes M, Alzubi S, Kanan T (2019) Ala al-fuqaha, and bilal hawashin. A survey on particle swarm optimization with emphasis on engineering and network applications Evolutionary Intelligence pp 1 – 17

41. Guo Z, Zhang L, Lu L, Bagheri M, Summers RM, Sonka M, Yao J (2018) Deep logismos: deep learning graph-based 3d segmentation of pancreatic tumors on ct scans. In: 2018 IEEE 15th International Symposium on Biomedical Imaging (ISBI 2018), IEEE, pp 1230 – 1233

42. Happ PN, Ferreira RS, Costa GA, Feitosa RQ, Bentes C, Gamba P (2015) Towards distributed region growing image segmentation based on mapreduce. In: 2015 IEEE International Geoscience and Remote Sensing Symposium (IGARSS), IEEE, pp 4352 – 4355

43. Hawashin B, Alzubi S, Kanan T, Mansour A (2019) An efficient semantic recommender method forarabic text. The Electronic Library

44. Hawashin B, Aqel D, Alzubi S, Elbes M (2019) Improving recommender systems using co-appearing and semantically correlated user interests. Recent Patents on Computer Science 12

45. Hawashin B, Aqel D, AlZu'bi S, Jararweh Y (2019) Novel weighted interest similarity measurement for recommender systems using rating timestamp. In: 2019 Sixth International Conference on Software Defined Systems (SDS), IEEE, pp 166 – 170

46. Hawashin B, Mansour A, Abukhait J, Khazalah F, AlZu'bi S, Kanan T, Obaidat M, Elbes M (2019) Efficient texture classification using independent component analysis. In: 2019 IEEE Jordan International Joint Conference on Electrical Engineering and Information Technology (JEEIT), IEEE, pp 544 – 547

47. Hawashin B, Alzubi S, Mughaid A, Fotouhi F, Abusukhon A (2020) An efficient cold start solution for recommender systems based on machine learning and user interests. In: 2020 Seventh International Conference on Software Defined Systems (SDS), IEEE, pp 220 – 225

48. Hussein WA, Ali BM, Rasid M, Hashim F (2017) Design and performance analysis of high reliability-optimal routing protocol for mobile wireless multimedia sensor networks. In: 2017 IEEE 13th Malaysia International Conference on Communications (MICC), IEEE, pp 136 – 140

49. (IEC) IEC, (NEMA) NEMA (2001) Nema standards publication no. nu2. National Electrical Manufacturers Association (NEMA) URL http://jrtassociates.com/pdfs/Pro-NM/20NEMA/20NU2.pdf

50. Jararweh Y, Alzubi S, Hariri S (2011) An optimal multi-processor allocation algorithm for high performance gpu accelerators. In: Applied Electrical Engineering and Computing Technologies (AEECT), 2011 IEEE Jordan Conference on, IEEE, pp 1 – 6

51. Jung C, Ke P, Sun Z, Gu A (2018) A fast deconvolution-based approach for single-image super-resolution with gpu acceleration. Journal of Real-Time Image Processing 14(2):501–512

52. Kanan T, Sadaqa O, Aldajeh A, Alshwabka H, AlZu'bi S, Elbes M, Hawashin B, Alia MA, et al (2019) A review of natural language pro- cessing and machine learning tools used to analyze arabic social media. In: 2019 IEEE Jordan International Joint Conference on Electrical Engi- neering and Information Technology (JEEIT), IEEE, pp 622–628

53. Kostrzewa M, Rathmann N, Kara K, Schoenberg SO, Diehl SJ (2015) Accuracy of percutaneous soft-tissue interventions using a multi-axis, c- arm ct system and 3d laser guidance. European Journal of Radiology 84(10):1970–1975, DOI https://doi.org/10.1016/j.ejrad.2015.06.028

54. Lafi M, Hawashin B, AlZu'bi S (2020) Maintenance requests labeling using machine learning classification. In: 2020 Seventh International Conference on Software Defined Systems (SDS), IEEE, pp 245–249

55. Li J, Wu Y, Zhao J, Lu K (2016) Multi-manifold sparse graph embedding for multi-modal image classification. Neurocomputing 173:501–510

56. Li J, Wu Y, Zhao J, Lu K (2017) Low-rank discriminant embedding for multiview learning. IEEE transactions on cybernetics 47(11):3516–3529

57. Liu W, Zhang T (2016) Multimedia hashing and networking. IEEE Mul- tiMedia 23(3):75–79

58. Martin M, Sciolla B, Sdika M, Wang X, Quetin P, Delachartre P (2018) Automatic segmentation of the cerebral ventricle in neonates using deep learning with 3d reconstructed freehand ultrasound imaging. In: 2018 IEEE International Ultrasonics Symposium (IUS), IEEE, pp 1–4

59. Mughaid A, Obeidat I, Hawashin B, AlZu'bi S, Aqel D (2019) A smart geo-location job recommender system based on social media posts. In: 2019 Sixth International Conference on Social Networks Analysis, Management and Security (SNAMS), IEEE, pp 505–510

60. Obeidat I, Mughaid A, Alzoubi S (2019) A secure encrypted protocol for clients' handshaking in the same network

61. Otake Y, Armand M, Armiger RS, Kutzer MD, Basafa E, Kazanzides P, Taylor RH (2012) Intraoperative image-based multiview 2d/3d registra- tion for image-guided orthopaedic surgery: incorporation of fiducial-based c-arm tracking and gpu-acceleration. IEEE transactions on medical imag- ing 31(4):948–962

62. Pawel B, Kawa J, Czajkowska J, Rudzki M, Pietka E (2011) Fuzzy con- nectedness in segmentation of medical images, a look at the pros and cons. International Conference on Fuzzy Computation Theory and Applications 2011:486–492

63. Peng H, Long F, Chi Z, Su W (2000) A hierarchical distributed genetic algorithm for image segmentation. In: Proceedings of the 2000 Congress on Evolutionary Computation. CEC00 (Cat. No. 00TH8512), IEEE, vol 1, pp 272–276

64. Ramadan R, Alqatawneh S, Ahalaiqa F, Abdel-Qader I, Aldahoud A, AlZoubi S (2019) The utilization of whatsapp to determine the obsessive-compulsive disorder (ocd): A preliminary study. In: 2019 Sixth Interna- tional Conference on Social Networks Analysis, Management and Security (SNAMS), IEEE, pp 561 – 564

65. Rezaee H, Aghagolzadeh A, Seyedarabi MH, Al Zu'bi S (2011) Tracking and occlusion handling in multi-sensor networks by particle filter. In: 2011 IEEE GCC Conference and Exhibition (GCC), IEEE, pp 397 – 400

66. Shadi A, Abbes A, et al. 3d medical volume segmentation using hybrid multiresolution statistical approaches. Advances in Artificial Intel- ligence 2010

67. Shih F (2009) Image Processing and Mathematical Morphology: Funda- mentals and Applications. CRC Press

68. Sim GH, Chang YC, Chuah TC (2011) Adaptive error protection for video transmission over ultra-wideband wireless multimedia sensor net- works. In: The International Conference on Information Networking 2011 (ICOIN2011), IEEE, pp 86 – 90

69. Sleman AA, Soliman A, Ghazal M, Sandhu H, Schaal S, Elmaghraby A, El- Baz A (2019) Retinal layers oct scans 3-d segmentation. In: 2019 IEEE In- ternational Conference on Imaging Systems and Techniques (IST), IEEE, pp 1 – 6

70. Sun Y, Tang J, Lei W, He D (2020) 3d segmentation of pulmonary nodules based on multi-view and semi-supervised. IEEE Access 8:26457 – 26467

71. Vazquez V, Eugenio M, Pedro G, Manlio FVC, Felipe AC, Eduardo AV, Claudio VS, Kirby GV, Manuel D, Javier P (2017) Assessment of intra- operative 3d imaging alternatives for ioert dose estimation. Zeitschrift fur Medizinische Physik 27(3):218 – 231

72. Wieclawek W, Pietka E (2007) Live-wire-based 3d segmentation method. In: 2007 29th Annual International Conference of the IEEE En- gineering in Medicine and Biology Society, pp 5645 – 5648, DOI 10.1109/IEMBS.2007.4353627

73. Wieclawek W, Pietka E (2015) Watershed based intelligent scissors. Computerized Medical Imaging and Graphics 43:122 – 129, DOI https://doi.org/10.1016/j.compmedimag.2015.01.003

74. Won HJ, Kim N, Kim GB, Seo JB, Kim H (2017) Validation of a ct-guided intervention robot for biopsy and radiofrequency ablation: experimental study with an abdominal phantom. Diagnostic and Interventional Radiol- ogy 23:233 – 237

75. Zhang L, Wang X, Yang D, Sanford T, Harmon S, Turkbey B, Wood BJ, Roth H, Myronenko A, Xu D, et al (2020) Generalizing deep learn- ing for medical image segmentation to unseen domains via deep stacked transformation. IEEE Transactions on Medical Imaging

76. Zhang X, Tan G, Chen M (2015) A reliable distributed convolutional neu- ral network for biology image segmentation. In: 2015 15th IEEE/ACM International Symposium on Cluster, Cloud and Grid Computing, IEEE, pp 777 – 780

77. Zhao N, Zheng X (2017) Multi-band blending of aerial images using gpu acceleration. In: Image and Signal Processing, BioMedical Engineering and

Informatics (CISP-BMEI), 2017 10th International Congress on, IEEE, pp 1 – 5

78. Zhou J, Qi J (2011) Fast and efficient fully 3d pet image reconstruction using sparse system matrix factorization with gpu acceleration. Physics in Medicine & Biology 56(20):6739